\documentclass[fp,twocolumn]{jpsj3}

\usepackage{amsmath}
\usepackage{amssymb}
\usepackage{bm}
\usepackage{graphicx}
\usepackage{times}
\usepackage{txfonts}

\usepackage[usenames]{color}

\newcommand{\siml}{\stackrel{<}{_\sim}}

\title{
Excitonic Phase Diagram of the Three-Chain Hubbard Model for Semiconducting and Semimetallic Ta$_2$NiSe$_5$
}
\author{
Kaoru {\sc Domon}$^1$\thanks{E-mail address: domon@phys.sc.niigata-u.ac.jp}, 
Takemi {\sc Yamada}$^2$, and 
Yoshiaki {\sc \=Ono}$^1$
}
\inst{
$^1$Department of Physics, Niigata University, Niigata, 950-2181, Japan\\
$^2$Department of Physics, Faculty of Science and Technology, Tokyo University of Science, Chiba, 278-8510, Japan
}
\abst{
Transition metal chalcogenide Ta$_2$NiSe$_5$, a promising material for the excitonic insulator, is investigated on the basis of the three-chain Hubbard model with two conduction ($c$) bands and one valence ($f$) band. 
In the semimetallic case where only one of two $c$ bands and the $f$ band cross the Fermi level, the transition from the $c$-$f$ compensated semimetal to the uniform excitonic order, the so-called excitonic insulator, takes place at low temperature as the same as in the semiconducting case. 
On the other hand, when another $c$ band also crosses the Fermi level, the system shows three types of Fulde-Ferrell-Larkin-Ovchinnikov (FFLO) excitonic orders characterized by the condensation of excitons with finite center-of-mass momentum $\bm{q}$ corresponding to the three types of nesting vectors between the imbalanced two $c$ and one $f$ Fermi surfaces. 
The obtained FFLO excitonic states are metallic in contrast to the excitonic insulator and are expected to be observed in the semimetallic Ta$_2$NiSe$_5$ under high pressure. 
The effect of the electron-lattice coupling is also discussed briefly and is found to induce the monoclinic distortion not only in the uniform excitonic state but also in the FFLO one resulting in the orthorhombic-monoclinic structural phase transition for both cases as observed in Ta$_2$NiSe$_5$ for both low-pressure semiconducting and high-pressure semimetallic regimes. 
}

\begin{document}
\maketitle

\section{Introduction}
\label{section1}
The quasi-one-dimensional (quasi-1-D) semiconductor Ta$_2$NiSe$_5$\cite{Sunshine1985} has received much attention as a strong candidate for the excitonic insulator (EI) which is characterized by the condensation of excitons and has been argued for about half a century\cite{Knox1963,Jerome1967,Halperin1968,Bronold2006,Ihle2008,Phan2010,Seki2011,Zenker2012,kunes2015}. 
It shows a structural transition from the orthorhombic to monoclinic phase at $T_c$=328 K\cite{DiSalvo1986}, below which the magnetic susceptibility shows a gradual drop and the flattening of the valence band top is observed in the ARPES experiment\cite{Wakisaka2009,wakisaka2012}. 
Several theoretical studies\cite{Kaneko2013,Seki2014,Sugimoto2016a,Sugimoto2016b,Matsuura2016} have revealed that the transition is well accounted for by the excitonic condensation from a normal semiconductor (orthorhombic) to the EI (monoclinic) from a mean-field analysis for the 1-D three-chain Hubbard model with electron-lattice coupling\cite{Kaneko2013,Sugimoto2016a} and from a variational cluster approximation for the extended Falicov-Kimball model\cite{Seki2014,Hamada2018}. 
Recent optical measurements are also consistent with the EI phase below $T_c$ \cite{Lu2017,Larkin2017}.

When the pressure is applied for Ta$_2$NiSe$_5$\cite{Nakano2018,Matsubayashi}, the structural phase transition temperature $T_c$ is suppressed and the system changes from semiconducting to semimetallic both above and below $T_c$, and then, $T_c$ finally becomes zero at a critical pressure $P_c\sim 8$GPa, around which the superconductivity is observed. 
Then, we have recently investigated the previous 1-D model also in the semimetallic case and have found that the difference of the band degeneracy, the two-fold degenerate conduction bands and the nondegenerate valence band, inevitably causes the imbalance of each Fermi wavenumber and results in a remarkable excitonic state characterized by the condensation of excitons with finite center-of-mass momentum $q$, the so-called Fulde-Ferrell-Larkin-Ovchinnikov (FFLO) excitonic state\cite{Yamada2016,Domon2016}, as previously discussed in the electron-hole bilayer systems with density imbalance\cite{Pieri2007,Yamashita2009,Zhu2010}.

Generally, the details of the Fermi surface (FS) are considered to be crucial to determine which types of excitonic orders are realized in the semimetallic case. As for Ta$_2$NiSe$_5$, two-fold degeneracy of the conduction bands, which is responsible for the FFLO excitonic state mentioned above, is known to be resolved by the inter-chain hoppings in a realistic present quasi-1-D model\cite{Sugimoto2016b}. 
To clarify the possible excitonic state in semimetallic Ta$_2$NiSe$_5$ under high pressure\cite{Nakano2018,Matsubayashi}, such effects of the inter-chain hoppings are important but has not been considered in the previous papers\cite{Yamada2016,Domon2016}. 
The purpose of this letter is to investigate the present quasi-1-D model with including the inter-chain hoppings, especially focusing on the semimetallic case in which the FFLO excitonic state is expected to be realized. 
We also briefly discuss the effect of the electron-lattice coupling which was found to account for the structural transition from the orthorhombic (normal) to monoclinic phase (EI) in the semiconducting case\cite{Kaneko2013} but has not been discussed in the semimetallic case.

\section{Model and Formulation}
\label{section2}
\begin{figure}[b]
\centering
\includegraphics[width=\linewidth]{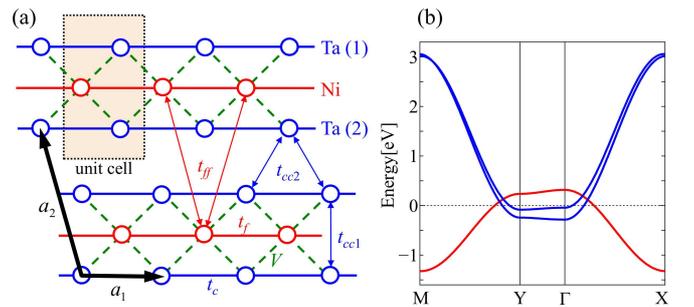}
\caption{
(Color online) 
(a) Schematic representation of the quasi-1-D three-chain Hubbard model for Ta$_2$NiSe$5$. $\bm{a}_{1}$ and $\bm{a}_{2}$ are unit lattice vectors. 
(b) Energy band structure in the semimetallic case of the normal state with $D=-0.4$ eV. 
}
\label{Fig1}
\end{figure}

The present quasi-1-D model for Ta$_2$NiSe$_5$\cite{Sugimoto2016b} is schematically shown in Fig. \ref{Fig1} (a), where a Ni chain and adjacent two Ta chains construct the 1-D three chain\cite{Kaneko2013} which is weakly coupled with the neighboring three chains, and is explicitly given by the Hamiltonian: $H=H_0+H'$ with
\begin{align}
H_0&=\sum_{\bm{k}\sigma}\sum_{\alpha=1,2}\epsilon_{\bm{k}}^{c}c^{\dagger}_{\bm{k}\alpha\sigma}c_{\bm{k}\alpha\sigma}+\sum_{\bm{k}\sigma}\epsilon_{\bm{k}}^{f}f^{\dagger}_{\bm{k}\sigma}f_{\bm{k}\sigma} \notag \\
    &+\sum_{\bm{k}\sigma}\left(\epsilon^{\prime}_{\bm{k}}c^{\dagger}_{\bm{k}1\sigma}c_{\bm{k}2\sigma}+\textrm{H.c.}\right), \label{eq:H0}\\
H'&=V\sum_{i\alpha}\sum_{\sigma\sigma^{\prime}}\left(c^{\dagger}_{\bar{\bm{R}_i}\alpha\sigma}c_{\bar{\bm{R}_i}\alpha\sigma}+c^{\dagger}_{\bm{R}_i\alpha\sigma}c_{\bm{R}_i\alpha\sigma} \right) f^{\dagger}_{\bm{R}_i\sigma^{\prime}}f_{\bm{R}_i\sigma^{\prime}},\label{eq:H'}
\end{align}
where $c_{\bm{k}\alpha\sigma}(c_{\bm{R}_i\alpha\sigma})$ and $f_{\bm{k}\sigma}(f_{\bm{R}_i\sigma})$ are the annihilation operators for Ta $5d$ ($c$) and Ni $3d$ ($f$) electrons with wavevector $\bm{k}$ (position $\bm{R}_i$ of the $i$-th unit cell and $\bar{\bm{R}_i}\equiv\bm{R}_i-\bm{a}_1$), spin $\sigma=\uparrow,\downarrow$, chain degrees of freedom for the $c$ electron $\alpha=1,2$. 
The noninteracting energies in Eq. (\ref{eq:H0}) are given by 
\begin{align}
\epsilon_{\bm{k}}^{c}&=\ D/2+2t_{c}\left(\cos{\bm{k}\cdot\bm{a}_{1}}-1\right)+|t_{cc1}+2t_{cc2}|, \\
\epsilon_{\bm{k}}^{f}&=-D/2+2t_{f}\left(\cos{\bm{k}\cdot\bm{a}_{1}}-1\right)\notag \\
&+2t_{ff}\left(\cos{\bm{k}\cdot(\bm{a}_{1}+\bm{a}_{2})}+\cos{\bm{k}\cdot\bm{a}_{2}}-2\right), \\
\epsilon^{\prime}_{\bm{k}}&= t_{cc1}+t_{cc2}(e^{i\bm{k}\cdot(\bm{a}_{1}+\bm{a}_{2})}+e^{i\bm{k}\cdot\bm{a}_{2}})
\end{align}
where the intra-chain hopping integrals along the Ta and Ni chains are set to $t_{c}=-0.80$ eV and $t_{f}=0.40$ eV, respectively, which are determined so as to fit the first-principles band structure of Ta$_2$NiSe$_5$\cite{Kaneko2013}, the inter-chain hopping integrals between the Ta chains are set to $t_{cc1}=-0.02$ eV and $t_{cc2}=-0.05$ eV which cause the splitting between the two-fold degenerate $c$ bands resulting in the bonding and antibonding $c$ bands as seen in Fig. \ref{Fig1} (b), and the inter-chain hopping integral between the Ni chains is set to $t_{ff}=0.01$ eV, respectively. 
As for the inter-chain hoppings, we choose the same values of Ref. \cite{Sugimoto2016b} except for $t_{cc2}=-0.05$ eV, which is slightly modified from that of Ref. \cite{Sugimoto2016b} ($t_{cc2}=-0.1$ eV), but the main conclusion obtained here is almost unchanged by this modification. 
$D$ is the energy gap between the $c$ and $f$ bands at the $\Gamma$ point, where $D>0$ ($D<0$) corresponds to the semiconducting (semimetallic) regime. 
As $D$ is considered to be a decreasing function of pressure, we vary $D$ as previously done in Refs. \cite{Yamada2016,Domon2016}. In Eq. (\ref{eq:H'}), we consider the intersite $c$-$f$ Coulomb interaction $V$ which is crucial for the excitonic order as shown below, while we neglect the on-site Coulomb interaction, which can be effectively included in $D$ and/or the chemical potential $\mu$ within the mean-field approximation by excluding the magnetic and density-wave-type orders\cite{Kaneko2013,Sugimoto2016a,Yamada2016,Domon2016,Sugimoto2016b}.

Now, we discuss the excitonic order within the mean-field approximation in which $H'$ in Eq. (\ref{eq:H'}) is replaced by $H'_\textrm{MF}=\sum_{\bm{q}} H'_\textrm{MF}(\bm{q})$ with
\begin{equation}
H'_\textrm{MF}(\bm{q})=\sum_{\bm{k}\sigma}\sum_{\alpha=1,2}\left(\Delta_{\bm{kq}\alpha}c_{\bm{k}\alpha\sigma}^{\dagger}f_{\bm{k}+\bm{q}\sigma}+\textrm{H.c.}\right)+\textrm{const}., 
\label{eq:MF}
\end{equation}
where the excitonic order parameter 
\begin{align}
\Delta_{\bm{kq}\alpha}=-\frac{V}{N}\sum_{\bm{k}^{\prime}}(1+e^{i(\bm{k}-\bm{k}^{\prime})\cdot\bm{a}_1})\langle f^{\dagger}_{\bm{k}^{\prime}+\bm{q}\sigma}c_{\bm{k}^{\prime}\alpha\sigma}\rangle \label{eq:OP1}
\end{align}
becomes finite when the condensation of excitonic $c$-$f$ pairs with center-of-mass momentum $\bm{q}$ takes place. 
We assume that the order parameter is independent of $\sigma$ but dependent on $\alpha$ in contrast to the case with the previous 1-D Hubbard model where it is independent of $\alpha$\cite{Yamada2016,Domon2016}. 
For the meanwhile, we consider the Fulde-Ferrell (FF) type state where $\Delta_{\bm{kq}\alpha}\ne 0$ for a specific $\bm{q}$, but we also discuss later the Larkin-Ovchinnikov (LO) type state where $\Delta_{\bm{kq}\alpha}\ne 0$ for both $\bm{q}$ and $-\bm{q}$.

Diagonalization of $H_\textrm{MF}(\bm{q})=H_0+H'_\textrm{MF}(\bm{q})$ yields the mean-field band dispersion $E_{\bm{kq}s}$ in the excitonic phase, where $s(=1,2,3)$ is the band index. 
In Eq. (\ref{eq:OP1}), $\Delta_{\bm{kq}\alpha}$ can be rewritten as 
\begin{equation}
\Delta_{\bm{kq}\alpha}=\Delta_{\bm{q}\alpha}^{(0)}+e^{i\bm{k}\cdot\bm{a}_1}\Delta_{\bm{q}\alpha}^{(1)}, 
\label{eq:OP2}
\end{equation}
where $\Delta_{\bm{q}\alpha}^{(0)}$ ($\Delta_{\bm{q}\alpha}^{(1)}$) represents the complex order parameter between the $f$-site and the $c$-site in the same (left neighboring) unit cell. 
Minimizing the free energy
\begin{align}
F({\bm{q}})= \sum_{\alpha\sigma}\frac{\left|\Delta_{\bm{q}\alpha}^{(0)}\right|^2+\left|\Delta_{\bm{q}\alpha}^{(1)}\right|^2}{V} 
-\frac{k_{B}T}{N}\sum_{\bm{k}s\sigma}\ln \left(1+e^{-\frac{E_{\bm{kq}s}-\mu}{k_{B}T}}\right)+\mu n 
\label{eq:FE}
\end{align}
with respect to $\Delta_{\bm{q}\alpha}^{(i)}$, we obtain the self-consistent equations to determine the order parameters: 
\begin{align}
\frac{V}{N}&\left(
   \begin{array}{cccc}
      \xi_{\bm{q}0}        & \xi_{\bm{q}1}         &  \eta_{\bm{q}0}     & \eta_{\bm{q}1}\\
      \xi_{\bm{q}1}^{*}   & \xi_{\bm{q}0}          &  \eta_{\bm{q}-1}   & \eta_{\bm{q}0}\\
      \eta_{\bm{q}0}^{*} & \eta_{\bm{q}-1}^{*} &  \xi_{\bm{q}0}       & \xi_{\bm{q}1}\\
      \eta_{\bm{q}1}^{*} & \eta_{\bm{q}0}^{*}   &  \xi_{\bm{q}1}^{*}  & \xi_{\bm{q}0}
    \end{array}
  \right)
  \left(
    \begin{array}{c}
      \Delta_{\bm{q}1}^{(0)} \\
      \Delta_{\bm{q}1}^{(1)} \\
      \Delta_{\bm{q}2}^{(0)} \\
      \Delta_{\bm{q}2}^{(1)}
    \end{array}
  \right)
=
  \left(
    \begin{array}{c}
      \Delta_{\bm{q}1}^{(0)} \\
      \Delta_{\bm{q}1}^{(1)} \\
      \Delta_{\bm{q}2}^{(0)} \\
      \Delta_{\bm{q}2}^{(1)}
    \end{array}
  \right) 
\label{eq:sce}
\end{align}
with
\begin{align}
\xi_{\bm{q}n}&=\sum_{\bm{k}}e^{i(\bm{k}\cdot\bm{a}_{1})n}u_{\bm{kq}}, \\
\eta_{\bm{q}n}&=\sum_{\bm{k}}e^{i(\bm{k}\cdot\bm{a}_{1})n}v_{\bm{kq}}, 
\end{align}
where 
\begin{align}
u_{\bm{kq}}&=-\sum_{s}\frac{E_{\bm{kq}s}-\epsilon_{\bm{k}}^{c}}{3(E_{\bm{kq}s}-E_{+})(E_{\bm{kq}s}-E_{-})}f(E_{\bm{kq}s}), \\
v_{\bm{kq}}&=-\sum_{s}\frac{\epsilon^{\prime}_{\bm{k}}}{3(E_{\bm{kq}s}-E_{+})(E_{\bm{kq}s}-E_{-})}f(E_{\bm{kq}s})
\label{eq:sce2}
\end{align}
with 
$f(E)=1/(1+e^{(E-\mu)/k_{B}T})$, 
$E_{\pm}=A\pm \sqrt{A^2+B/3}$, 
$A=(2\epsilon_{\bm{k}}^{c}+\epsilon_{\bm{k}+\bm{q}}^{f})/3$ and 
$B=2\epsilon_{\bm{k}}^{c}\epsilon_{\bm{k}+\bm{q}}^{f}+{\epsilon_{\bm{k}}^{c}}^{2}-\sum_{\alpha}\left|\Delta_{\bm{kq}\alpha}\right|^{2}-{\left|\epsilon_{\bm{k}}^{\prime}\right|}^{2}$. 
In the above, the chemical potential $\mu$ is determined so as to fix the number of electrons per unit cell to $n=n_c+n_f$. 
Generally, Eqs. (\ref{eq:sce})-(\ref{eq:sce2}) yield the self-consistent solutions of $\Delta_{\bm{q}\alpha}^{(0)}$ and $\Delta_{\bm{q}\alpha}^{(1)}$ for various values of $\bm{q}$. 
Therefore, we finally determine the stable solution by minimizing the free energy in Eq. (\ref{eq:FE}) with respect to $\bm{q}$. 
It is noted that, when we set $\epsilon^{\prime}_{\bm{k}}=0$ and neglect the $\alpha$-dependence of the order parameters, the above self-consistent equations coincide with those in the previous 1-D Hubbard model\cite{Yamada2016,Domon2016}.

By solving Eqs. (\ref{eq:sce})-(\ref{eq:sce2}) with Eqs. (\ref{eq:MF})-(\ref{eq:FE}) numerically, we obtain the four complex order parameters $\Delta_{\bm{q}\alpha}^{(i)}$ with $i=0,1$ and $\alpha=1,2$ and find that those amplitudes are independent of $i$ and $\alpha$ while those phases depend on $i$ and $\alpha$. 
Then, we further rewrite $\Delta_{\bm{kq}\alpha}$ in Eq. (\ref{eq:OP2}) as 
\begin{align}
\Delta_{\bm{kq}\alpha} =\Delta_{\bm{q}}\left(1+e^{i\bm{k}\cdot\bm{a}_1}e^{-i\phi_{\bm{q}}}\right)e^{-i(\alpha-1)\psi_{\bm{q}}}, 
\end{align}
where $\Delta_{\bm{q}}$ is the amplitude of the order parameter, $\phi_{\bm{q}}(\psi_{\bm{q}})$ is the relative phase between the nearest neighbor $c$-$f$ pair with $c$-site to the right (upper) side of $f$-site and that to the left (lower) side.

\begin{table}[t]
\centering
\vspace{+0.3cm}
\begin{tabular}{c|ccccc|c}
\hline
Excitonic phase & $q_x$ & $q_y$ & $\phi_{\bm{q}}$ & $\psi_{\bm{q}}$ & M/I & No. of e/h FSs  \\ \hline
Uniform & 0 & 0 & 0 & 0 & I & 0/0 or 1/1  \\
FFLO1   & $\ne 0$ & 0     & $\ne 0$ & 0 & M & 2/1 \\
FFLO2   & $\ne 0$ & $\pi$ & $\ne 0$ & $\pi$ & M & 2/1 \\
FFLO3   & 0 & $\pi$ & 0 & 0 & M & 2/1 \\ 
\hline
\end{tabular}
\vspace{-0.0cm}\caption{
List of the obtained excitonic phases with their characterized parameters: the center-of-mass momentum $\bm{q}=(q_x,q_y)$, the relative phases of the order parameter $\phi_{\bm{q}}$ and $\psi_{\bm{q}}$, and the metallic (M) or insulating (I) ground state. The last column shows the number of electron/hole FSs of the normal states which have the corresponding $c$-$f$ nesting yielding the excitonic states in each row. 
}
\label{table1}
\end{table}

\begin{figure}[t]
\centering
\vspace{+0.3cm}
\includegraphics[width=\linewidth]{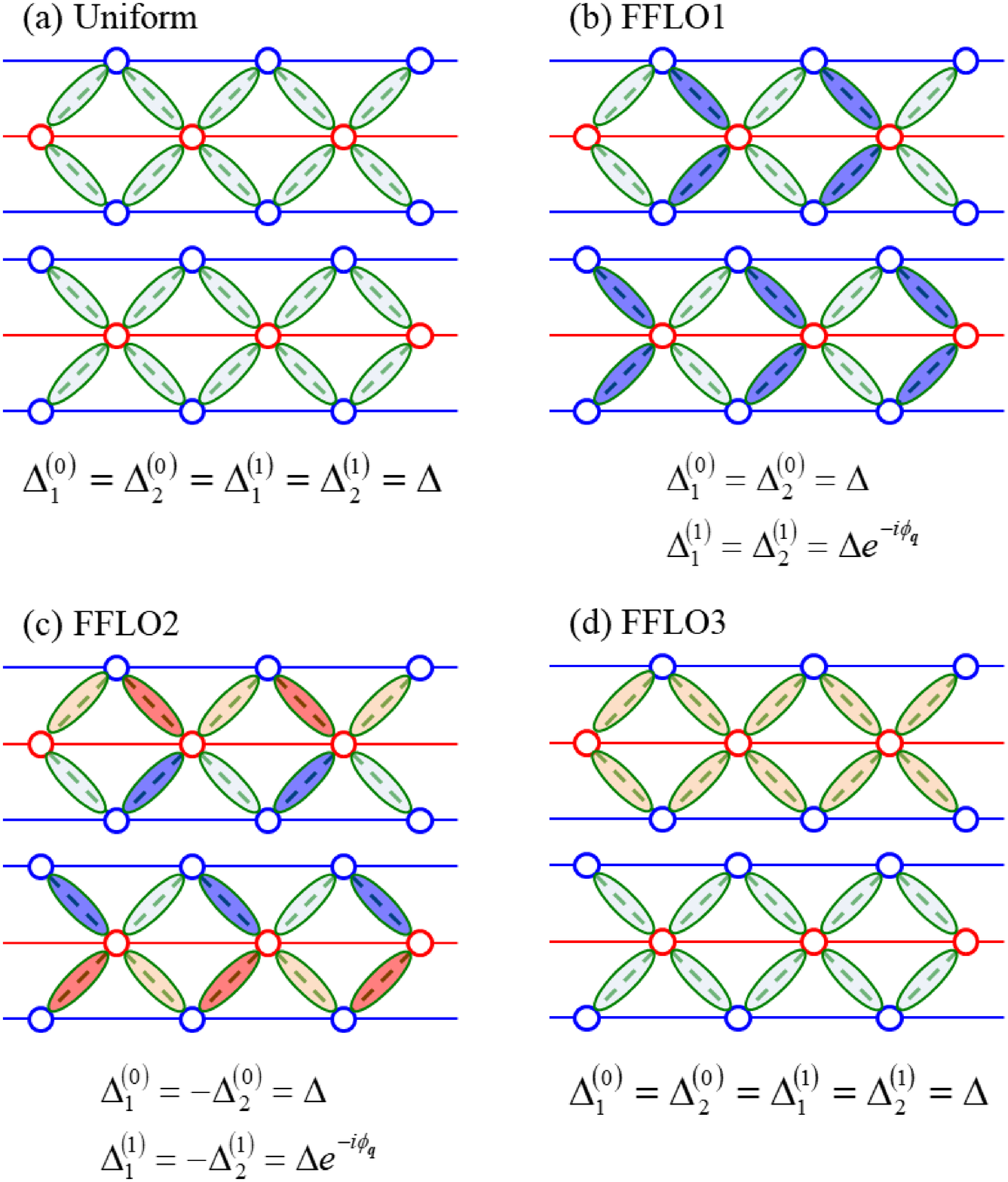}
\caption{
(Color online) 
Schematic representation of the order parameters in the real space for the four types of excitonic phases: uniform (a), FFLO1 (b), FFLO2 (c) and FFLO3 (d), as listed in Table \ref{table1}. 
The color of the ellipses between Ta and Ni means the phase of the order parameters.
}
\label{Fig2}
\end{figure}

We obtain four distinct excitonic phases: uniform, FFLO1, FFLO2 and FFLO3 phases, as summarized in Table \ref{table1} and schematically shown in Fig. \ref{Fig2} where the configurations of the order parameters in the real space are plotted. 
The uniform and the FFLO states with $q_{y}=0$, where the order parameter is independent of chain as shown in Figs. \ref{Fig2} (a) and (b), are essentially the same as those obtained in the previous 1-D Hubbard model\cite{Yamada2016,Domon2016}, but the FFLO2 and the FFLO3 states with $q_{y}=\pi$, where the order parameter changes its sign between the adjacent  three-chains as shown in Figs. \ref{Fig2} (c) and (d), are the novel states specific in the present quasi-1-D Hubbard model as explicitly discussed in Sec. \ref{section3}. 
We note that, in the FFLO states, the excitonic order parameters are spatially oscillating as shown in Figs. \ref{Fig2} (b)-(d) but the electron density is uniform without any density wave. 
In the present study, we set $n=2$ and $V=0.6$ eV, and vary $T$ and $D$ as parameters. 
Here and hereafter, the energy is measured in units of eV.

\section{Results}
\label{section3}
\subsection{$D$-dependence of excitonic order parameters}
\label{subsection31}
\begin{figure}[t]
\centering
\vspace{+0.3cm}
\includegraphics[width=6.6cm]{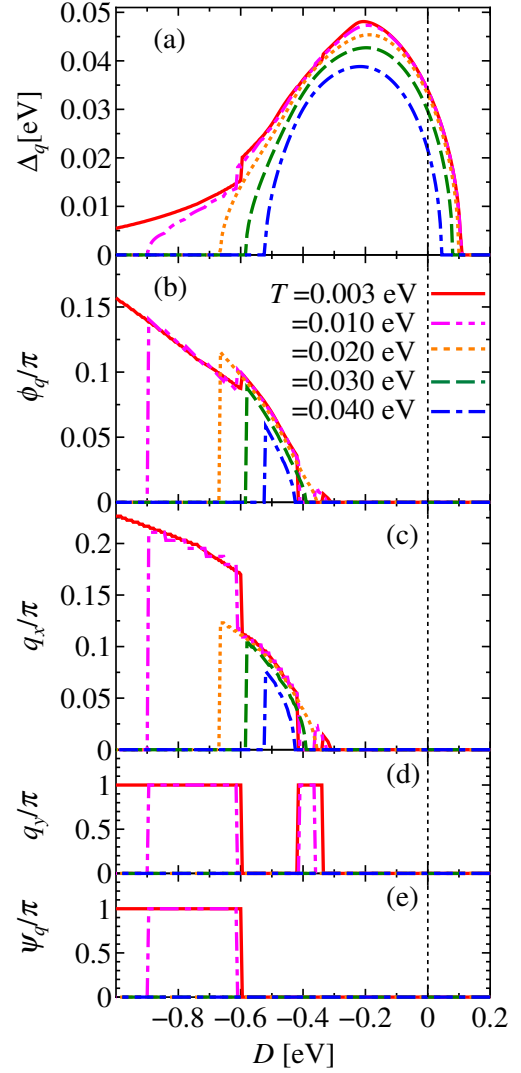}
\caption{(Color online) 
The amplitude of the excitonic order parameter $\Delta_{\bm{q}}$ (a), its relative phases $\phi_{\bm{q}}$ (b) and $\psi_{\bm{q}}$ (e) and the center-of-mass momentum $\bm{q}=(q_x,q_y)$ (c) and (d) as functions of the energy gap $D$ for several values of temperature $T$ for $n$=2 and $V$=0.6. 
}
\label{Fig3}
\end{figure}
In Figs. \ref{Fig3} (a)-(e), we plot $\Delta_{\bm{q}}$, $\phi_{\bm{q}}$, $q_{x}$, $q_{y}$ and $\psi_{\bm{q}}$ as functions of $D$ for several values of $T$. 
At a relatively high temperature $T=0.040$, the excitonic phase with $\Delta_{\bm{q}}\ne 0$ is observed for $-0.52<D<0.04$. 
When the corresponding normal state obtained by assuming $\Delta_{\bm{q}}= 0$ is a narrow gap semiconductor ($D>0$) or a semimetal in which only one of two $c$ bands and the one $f$ band cross the Fermi level ($-0.4\siml D<0$), the uniform excitonic state with $q_{x}=q_{y}=\phi_{\bm{q}}=\psi_{\bm{q}}=0$ takes place. 
On the other hand, when another conduction band also crosses the Fermi level in the corresponding normal state  ($D\siml -0.4$), the FFLO1 state with $q_{x}\ne 0$, $q_{y}=0$, $\phi_{\bm{q}}\ne 0$ and $\psi_{\bm{q}}=0$ takes place. 
With decreasing $T$, the region of the excitonic phase increases. 
In addition, at lower temperatures $T=0.010$ and $T=0.003$, we observe the other types of FFLO states: the FFLO2 state with $q_{x}\ne 0$, $q_{y}=\pi$, $\phi_{\bm{q}}\ne 0$ and $\psi_{\bm{q}}=\pi$ for large $|D|$ and the FFLO3 with $q_{x}= 0$, $q_{y}=\pi$ and $\phi_{\bm{q}}=\psi_{\bm{q}}=0$ for small $|D|$. 

Detailed analysis of the free energy reveal that the phase transitions from the FFLO1 to the FFLO2 and from the FFLO1 to the FFLO3 are the first-order ones, while the other phase transitions from the normal to the excitonic states (uniform, FFLO1 and FFLO2) and from the uniform to the FFLO1 are the second-order ones, which are also observed when $T$ is varied as shown in Sec. \ref{subsection32}.

\subsection{$T$-dependence of excitonic order parameters}
\label{subsection32}
\begin{figure}[t]
\centering
\vspace{+0.3cm}
\includegraphics[width=6.3cm]{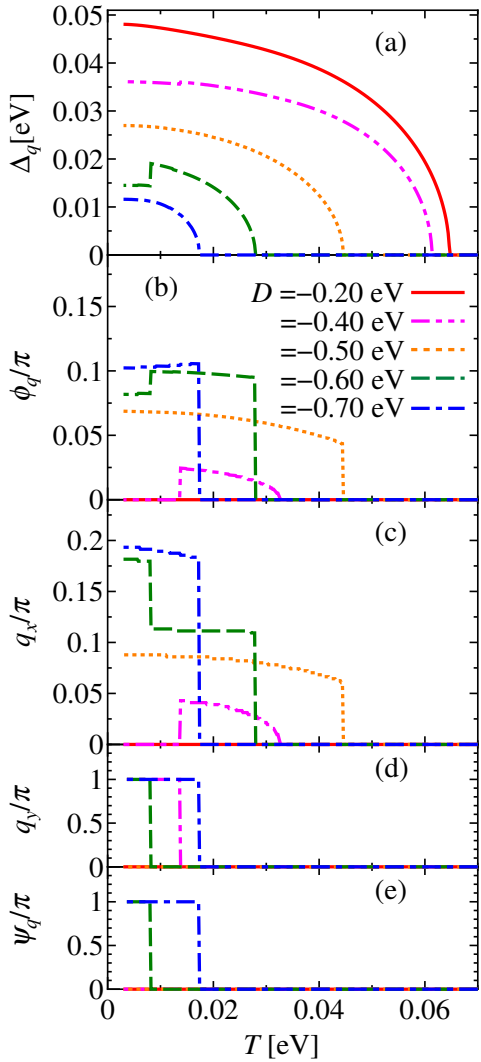}
\caption{(Color online) 
The amplitude of the excitonic order parameter $\Delta_{\bm{q}}$ (a), its relative phases $\phi_{\bm{q}}$ (b) and $\psi_{\bm{q}}$ (e) and the center-of-mass momentum $\bm{q}=(q_x,q_y)$ (c) and (d) as functions of temperature $T$ for several values of the energy gap $D$ for $n=2$ and $V=0.6$. 
}
\label{Fig4}
\end{figure}
In Figs. \ref{Fig4} (a)-(e), we plot $\Delta_{\bm{q}}$, $\phi_{\bm{q}}$, $q_{x}$, $q_{y}$ and $\psi_{\bm{q}}$ as functions of $T$ for several values of $D$.
In the semimetallic case with a small band overlapping $D=-0.2$, the second-order phase transition from the normal semimetal to the uniform excitonic state (EI) takes place below which the order parameter $\Delta_{\bm{q}}$ with $q_{x}=q_{y}=0$ monotonically increases as similar to the semiconducting case $D>0$ (not shown). 
When the band overlapping is relatively large $D=-0.5$, we observe the second-order phase transition from the normal semimetal to the FFLO1 as previously observed in the 1-D Hubbard model\cite{Yamada2016,Domon2016}. 
For a further large band overlapping $D=-0.7$, the system shows the second-order phase transition from the normal semimetal to the FFLO2 which is a specific feature in the present quasi-1-D Hubbard model. 
In addition to the above mentioned cases where the single phase transition takes place, we observe the consequent phase transitions in the intermediate parameter regimes: the normal-uniform-FFLO1-FFLO3 transitions for $D=-0.4$ and the normal-FFLO1-FFLO2 transitions for $D=-0.6$ (see also Fig. \ref{Fig5}). 

\subsection{Excitonic phase diagram on $D$-$T$ plane}
\label{subsection33}
\begin{figure}[t]
\centering
\vspace{+0.3cm}
\includegraphics[width=6.7cm]{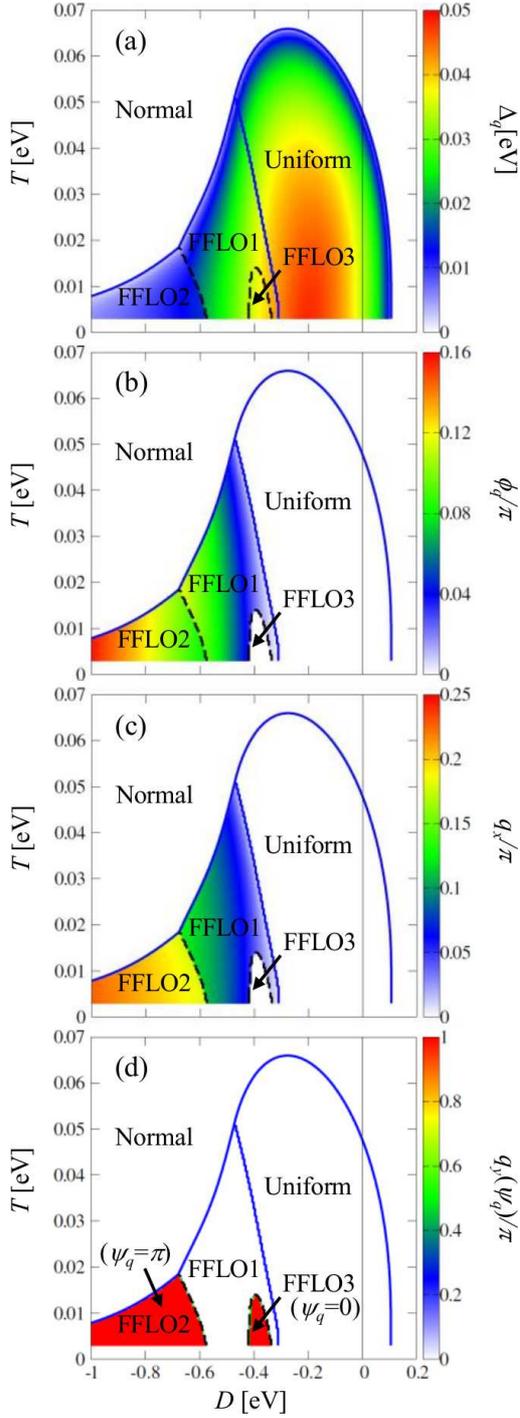}
\caption{
(Color online) 
Excitonic phase diagrams of the present quasi-1-D Hubbard model for Ta$_2$NiSe$_5$ in $D$-$T$ plane for $n=2$ and $V=0.6$ on the color maps of $\Delta_{\bm{q}}$ (a),  $\phi_{\bm{q}}$ (b), $q_x$ (c), and $q_y$ with $\psi_{\bm{q}}$ (d). Solid and dashed lines indicate the second- and the first-order phase transitions, respectively. 
}
\label{Fig5}
\end{figure}
Systematic calculations for various values of $D$ and $T$ yield  the excitonic phase diagram on the $D$-$T$ plane as shown in Figs. \ref{Fig5} (a)-(d) where $\Delta_{\bm{q}}$, $\phi_{\bm{q}}$, $q_{x}$ and $q_{y}$ with $\psi_{\bm{q}}$ are plotted. We find that the excitonic order with $\Delta_{\bm{q}} \ne 0$ is realized for $D\siml 0.1$ below the transition temperature $T_c$. 
In the case with a narrow gap semiconductor for $0<D\siml 0.1$, the transition from the semiconductor to the uniform excitonic state characterized by the BEC of excitons takes place\cite{Kaneko2013}. 
When the gap $D$ decreases, $T_c$ rapidly increases with increasing carrier density as expected in the BEC regime. 
In the semimetallic case, $T_c$ still increases with decreasing $D$ for $-0.3\siml D<0$ and shows a maximum at $D\sim -0.3$, and then, monotonically decreases for $D\siml -0.3$ where the transition from the semimetal to the BCS-like excitonic condensation takes place. 

When the normal state above $T_c$ is semiconducting ($D>0$) or the semimetallic with a slight band overlapping ($-0.47 \siml D<0 $), the transition from the normal to the uniform excitonic order takes place. For the case with a relatively large band overlapping $-0.68\siml D\siml -0.47$, we observe the transition from the normal semimetal to the FFLO1 excitonic order. Then, the system shows the transition from the normal semimetal to the FFLO2 excitonic order for larger band overlapping $D \siml -0.68$. We also observe the FFLO3 order at low temperature in a narrow region with $-0.42\siml D\siml -0.33$. The phase transitions from the FFLO1 to the FFLO2 and from the FFLO1 to the FFLO3 are the first-order ones, while the other phase transitions from the normal to the excitonic states (uniform, FFLO1 and FFLO2) and from the uniform to the FFLO1 are the second-order ones as mentioned before. The details of the energy bands and the FSs are considered to be crucial to determine which types of excitonic orders are realized and are shown in Secs. \ref{subsection34} and \ref{subsection35}.

\subsection{Energy bands}
\label{subsection34}
\begin{figure}[t]
\centering
\vspace{+0.3cm}
\includegraphics[width=8.5cm]{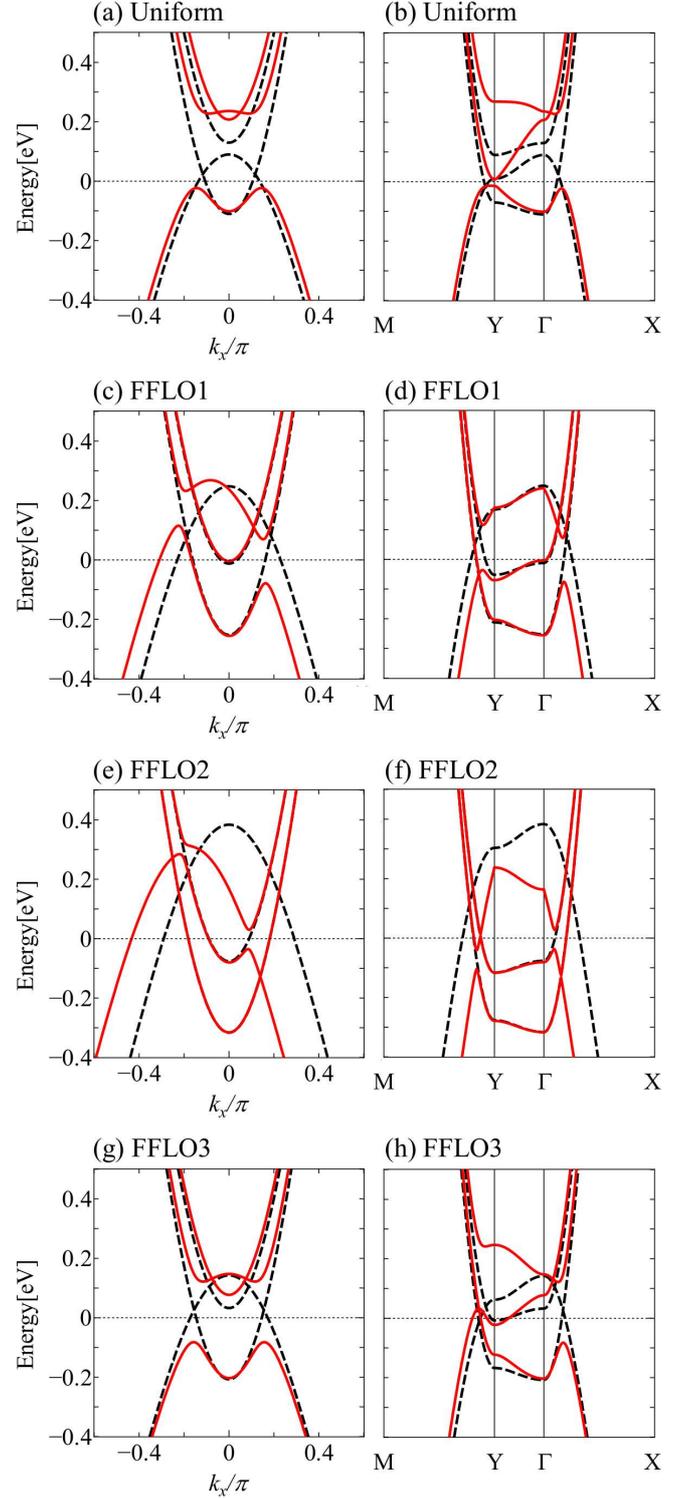}
\caption{(Color online) 
Energy bands in the excitonic states (solid lines) for $D=-0.2$ (uniform) (a) and (b), $D=-0.5$ (FFLO1) (c) and (d), $D=-0.7$ (FFLO2) (e) and (f), $D=-0.35$ (FFLO3) (g) and (h), along the $k_x$-axis with $k_y=0$ and along the M-Y-$\Gamma$-X (see Fig. \ref{Fig7}), respectively, for $n=2$, $V=0.6$ and $T=0.003$, together with those in the corresponding normal states with assuming $\Delta_{\bm{q}}=0$ (dashed lines). 
}
\label{Fig6}
\end{figure}
In Figs. \ref{Fig6} (a)-(h), we plot the energy bands for four typical values of $D=-0.2$, $-0.5$, $-0.7$ and $-0.35$ in which the uniform, FFLO1 FFLO2 and FFLO3 excitonic orders take place, respectively. 
When only one of two $c$ bands, the bonding $c$ band, and the one $f$ band cross the Fermi level, the transition from the normal semimetal to the uniform excitonic state occurs as shown in Figs. \ref{Fig6} (a) and (b), as similar as in the semiconducting case (not shown). 
On the other hand, when another $c$ band, the antibonding $c$ band, also crosses the Fermi level, the system shows the transition from the normal semimetal to the FFLO1 excitonic state for $D=-0.5$ as shown in Figs. \ref{Fig6} (c) and (d), and that to the FFLO2 one for $D=-0.7$ as shown in Figs. \ref{Fig6} (e) and (f). In addition, we also observe the FFLO3 excitonic state at low temperature for $D=-0.35$ as shown in Figs. \ref{Fig6} (g) and (h).

We note that the uniform excitonic state is insulating with a finite energy gap (EI) as seen in Fig. \ref{Fig6} (b) with the energy gap $\sim 0.02$, except for the narrow region in the vicinity of the phase boundary for the FFLO1 phase with $-0.31\siml D\siml -0.26$. On the other hand, the FFLO excitonic states (FFLO1, FFLO2 and FFLO3) are metallic as seen in Figs. (d), (f) and (h), and are expected to be observed in Ta$_2$NiSe$_5$ under high pressure where the system is found to be semimetallic both above and below $T_c$ of the orthorhombic-monoclinic structural phase transition\cite{Nakano2018,Matsubayashi}. 
In the FFLO1 and FFLO2 states, the band dispersions are asymmetric with respect to $k_x=0$ as seen in Figs. \ref{Fig6} (c) and (e) as the same as in the previous 1-D Hubbard model\cite{Yamada2016,Domon2016}. 
It is noted that the asymmetric band dispersion is a specific feature of the FF type state where $\Delta_{\bm{q}}\ne 0$ only for a single-$\bm{q}$, while the band dispersion is symmetric in the LO type state where $\Delta_{\bm{q}}=\Delta_{-\bm{q}}\ne 0$ (see Fig. \ref{Fig9}).

\subsection{Fermi surfaces}
\label{subsection35}
\begin{figure}[t]
\centering
\vspace{+0.0cm}
\includegraphics[width=4.8cm]{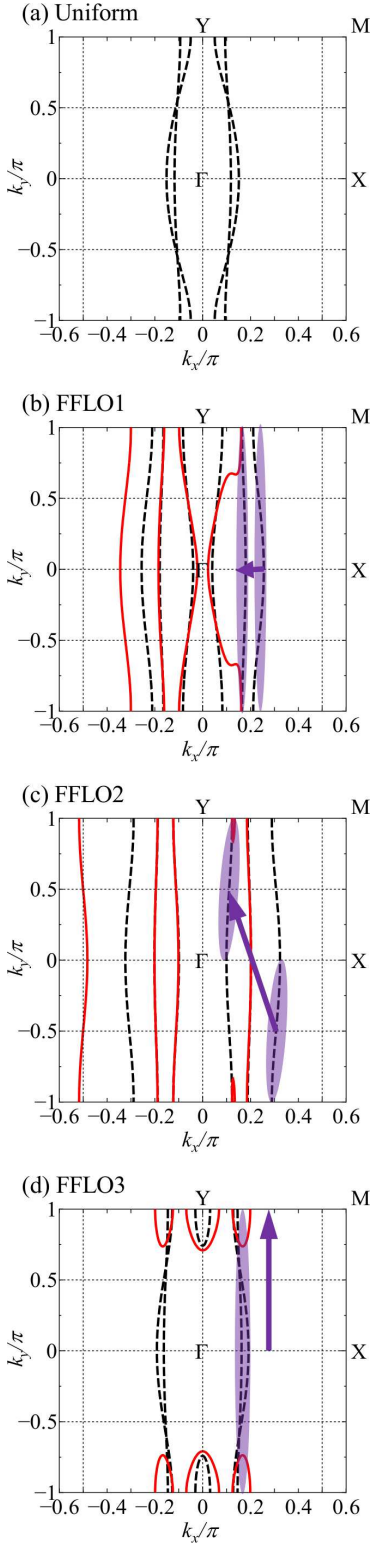}
\caption{(Color online) 
Fermi surfaces in the excitonic states (solid lines) for $D=-0.2$ (uniform) (a), $D=-0.5$ (FFLO1) (b), $D=-0.7$ (FFLO2) (c) and $D=-0.35$ (FFLO3) (d), respectively, for $n=2$, $V=0.6$ and $T=0.003$, together with those in the corresponding normal states (dashed lines) where arrows indicate the nesting vectors between the $c$ and $f$ FSs. 
}
\label{Fig7}
\end{figure}

Figures \ref{Fig7} (a)-(d) show the FSs for four typical values of $D=-0.2$, $-0.5$, $-0.7$ and $-0.35$ in which the uniform, FFLO1 FFLO2 and FFLO3 excitonic orders are realized, respectively. In the case with a small band overlapping $D=-0.2$, only one of two $c$ bands and the $f$ band cross the Fermi level, and then, the $c$ and $f$ FSs are compensated as shown in Figs. \ref{Fig6} (a) and (b). In this case, a good nesting between the $c$ and the $f$ FSs with the nesting vector $\bm{q}=0$ (see Fig. \ref{Fig7} (a)) results in the uniform excitonic order with a finite energy gap as the same as in the semiconducting case with $D>0$ (not shown)\cite{Kaneko2013}.

When the band overlapping is relatively large $D=-0.5$, the antibonding $c$ band also crosses the Fermi level in addition to the bonding $c$ band and the $f$ band as shown in Figs. \ref{Fig6} (a) and (b), and then, the imbalance of the number of FSs, two $c$ and one $f$, inevitably causes the nonzero nesting vector $\bm{q}=(q_x,0)$ between the bonding $c$ and the $f$ FSs (see Fig. \ref{Fig7} (b)) resulting in the FFLO1 excitonic order with $q_x\ne 0$. When the band overlapping becomes further large $D=-0.7$, the nesting effect between the antibonding $c$ and the $f$ FSs with the nesting vector $\bm{q}=(q_x,\pi)$ (see Fig. \ref{Fig7} (c)) dominates over that between the bonding $c$ and the $f$ FSs mentioned above and results in the FFLO2 excitonic order. In addition, in a narrow region around $D= -0.35$, another nesting effect between the bonding $c$ and the $f$ FSs with the nesting vector $\bm{q}=(0,\pi)$ (see Fig. \ref{Fig7} (d)) dominates over the other nesting effects, and then results in the FFLO3 excitonic order. 
We again find that the FFLO excitonic states (FFLO1, FFLO2 and FFLO3) are metallic as seen in Figs. \ref{Fig7} (b), (c) and (d) in contrast to the uniform excitonic state (see Fig. \ref{Fig7} (a)) and are expected to be observed in Ta$_2$NiSe$_5$ under high pressure where the system is semimetallic both above and below $T_c$\cite{Nakano2018,Matsubayashi}.

\subsection{Effect of electron-lattice coupling}
\label{subsection36}
Finally, we discuss the effect of the electron-lattice coupling responsible for the orthorhombic-monoclinic structural phase transition. As the monoclinic distortion yields the $c$-$f$ hybridization which is absent in the orthorhombic phase (see Eq. (\ref{eq:H0})), we consider the electron-lattice interaction Hamiltonian together with the elastic energy as follows
\begin{equation}
H_{\rm ep}=g\sum_{\bm{k}\sigma}\sum_{\alpha=1,2}(\delta_{\alpha}c_{\bm{k}\alpha\sigma}^{\dagger}f_{\bm{k}\sigma}+\textrm{H.c.})+\frac{K}{2}\sum_{\alpha=1,2}\delta_{\alpha}^2
\end{equation}
where $g$ is the electron-lattice coupling constant, $K$ is the string constant and $\delta_1=-\delta_2$ is the uniform shear distortion of the chain corresponding to the monoclinic phase\cite{Kaneko2013}. With including $H_{\rm ep}$ in the mean-field Hamiltonian Eq. (\ref{eq:MF}), the excitonic order parameter Eq. (\ref{eq:OP1}) is replaced as 
$\Delta_{\bm{kq}\alpha}\to \Delta_{\bm{kq}\alpha}+g\delta_{\alpha}\delta_{\bm{q},0}$. Then, the uniform excitonic order $\Delta_{\bm{q}=0}$  induces the monoclinic distortion $\delta_{\alpha}$ due to the effect of $g$ resulting in the orthorhombic-monoclinic structural phase transition\cite{Kaneko2013}.

To discuss the monoclinic distortion $\delta_{\alpha}$, which is induced by the uniform excitonic order $\Delta_{\bm{q}=0}$ as mentioned above, in the FFLO excitonic state, we consider the LO type state with $\Delta_{\bm{q}}=\Delta_{-\bm{q}}\ne 0$ where $\Delta_{\bm{q}=0}$ is also determined self-consistently and then generally becomes finite, instead of the FF type state with $\Delta_{\bm{q}}\ne 0$ and $\Delta_{-\bm{q}}=\Delta_{\bm{q}=0}=0$ considered in the above subsections. As explicit numerical calculations for the LO type states in the present quasi-1-D Hubbard model are rather complicated, we here employ the previous 1-D three-chain Hubbard model\cite{Yamada2016,Domon2016} and consider the simplest LO state where $\Delta_{q=\pi}=\Delta_{q=-\pi}$ and $\Delta_{q=0}$ are finite. We note that the transition temperature $T_c$ from the normal to the LO state is the same as that to the FF state within the mean-field approximation but the free energy of the LO state is always lower than that of the FF state in the both cases with and without the electron-lattice coupling $g$. We also note that the electron density is uniform without any density wave in the present LO state.

\begin{figure}[t]
\centering
\vspace{+0.3cm}
\includegraphics[width=7.5cm]{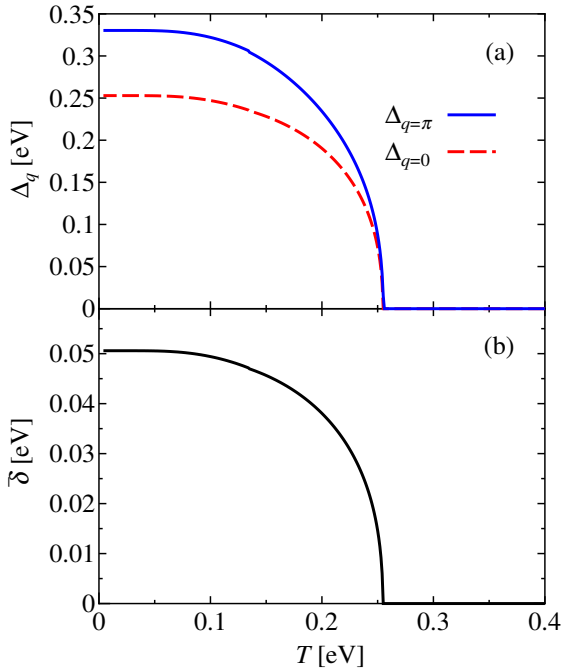}
\caption{(Color online) 
Temperature dependence of the amplitudes of the LO type excitonic order parameters $\Delta_{q=\pi}$ (solid line) and $\Delta_{q=0}$ (dashed line) (a) and the uniform lattice distortion $\bar{\delta}$ (b) for $V=2$, $V_\textrm{ep}=0.1$ and $D=-2.4$ in the 1-D three-chain Hubbard model. 
}
\label{Fig8}
\end{figure}

In Fig. \ref{Fig8} (a), the amplitudes of the LO type excitonic order parameters $\Delta_{q=\pi}(=\Delta_{q=-\pi})$ and $\Delta_{q=0}$ are plotted as functions of $T$ for $V=2$, $V_\textrm{ep}\equiv g^2/K=0.1$ and $D=-2.4$ in the 1-D three-chain Hubbard model. The second-order phase transition from the normal to the LO state takes place at $T_c \sim 0.26$ below which both $\Delta_{q=\pi}$ and $\Delta_{q=0}$ are finite. Then, the uniform lattice distortion corresponding to the monoclinic phase, $\bar{\delta}\equiv g\delta_1=-g\delta_2$, is induced by $\Delta_{q=0}$ due to the effect of $g$ resulting in the second-order orthorhombic-monoclinic structural phase transition at $T_c$ as shown in Fig. \ref{Fig8} (b), as similar in the case with the uniform excitonic state\cite{Kaneko2013}. 

\begin{figure}[t]
\centering
\vspace{+0.3cm}
\includegraphics[width=7.0cm]{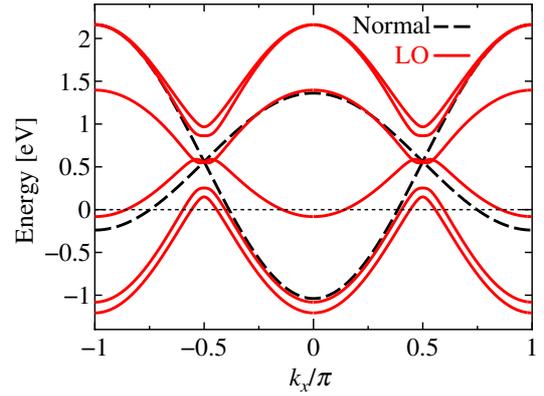}
\caption{(Color online) 
Energy bands in the LO type excitonic state (solid lines) together with those in the corresponding normal state with assuming $\Delta_{q}=0$ (dashed lines) for $V=2$, $V_\textrm{ep}=0.1$ and $D=-2.4$ in the 1-D three-chain Hubbard model. 
}
\label{Fig9}
\end{figure}

Figure \ref{Fig9} shows the energy bands in the LO type excitonic state for $V=2$, $V_\textrm{ep}=0.1$ and $D=-2.4$ in the 1-D three-chain Hubbard model. We find that the band dispersions are symmetric with respect to $k_x=0$ in contrast to the FF type excitonic states shown in Figs. \ref{Fig6} (c) and (e). 
We note that the LO excitonic state is semimetallic as the same as the FF excitonic states (see Figs. \ref{Fig6} (c)-(h)), in contrast to the uniform excitonic state (see Figs. \ref{Fig6} (a) and (b)), and are expected to be observed in Ta$_2$NiSe$_5$ under high pressure where the system is semimetallic both above and below the orthorhombic-monoclinic structural transition temperature $T_c$\cite{Nakano2018,Matsubayashi}.

\section{Summary and Discussions}
\label{section4}
In summary, we have investigated the quasi-1-D three-chain Hubbard model for Ta$_2$NiSe$_5$ in the presence of the inter-chain hoppings which cause the splitting between the two-fold degenerate $c$ bands resulting in the bonding and antibonding $c$ bands. In the semimetallic case with a small band overlapping where only the bonding $c$ bands and the $f$ band cross the Fermi level, the transition from the $c$-$f$ compensated semimetal to the uniform excitonic state corresponding to the EI takes place as the same as in the semiconducting case. On the other hand, when the antibonding $c$ band also crosses the Fermi level in addition to the bonding $c$ and the $f$ band, the system shows three types of FFLO excitonic states characterized by the condensation of excitons with finite center-of-mass momentum $\bm{q}$ as follows: FFLO1 with $\bm{q}=(q_x,0)$ corresponding to the nesting vector between the bonding $c$ and the $f$ FSs, FFLO2 with $\bm{q}=(q_x,\pi)$ between the antibonding $c$ and the $f$ FSs, and FFLO3 with $\bm{q}=(0,\pi)$ between the bonding $c$ and the $f$ FSs. The uniform and the FFLO1 states with $q_{y}=0$ are essentially the same as those in the previous 1-D model\cite{Yamada2016,Domon2016}, but the FFLO2 and the FFLO3 states with $q_{y}=\pi$ are the novel states specific in the quasi-1-D model. 
The obtained FFLO states here are all semimetallic which is consistent with the high-pressured situation of Ta$_2$NiSe$_5$, while the conventional uniform excitonic state with shallow $|D|$ region is fully insulating. Hence such the FFLO excitonic states are expected to be observed in semimetallic Ta$_2$NiSe$_5$ under high pressure\cite{Nakano2018,Matsubayashi}.

We have also briefly discussed the effect of the electron-lattice coupling responsible for the orthorhombic-monoclinic structural phase transition. Considering the LO type state where not only the FFLO excitonic order parameters with $\pm\bm{q}$ but also the uniform one with $\bm{q}=0$ become finite, the monoclinic distortion is found to be induced in the FFLO excitonic state resulting in the orthorhombic-monoclinic structural phase transition as the same as in case with the uniform excitonic state\cite{Kaneko2013,Sugimoto2016a}. 
This seems to be consistent with the experimental observation in Ta$_2$NiSe$_5$ where the orthorhombic-monoclinic structural phase transition is observed not only in the semiconducting case at ambient and low pressure but also in the semimetallic case at high pressure\cite{Nakano2018,Matsubayashi}.

In comparison with the actual material Ta$_2$NiSe$_5$, however, the present theory is still incomplete and we need to further consider several effects in the following: 
(1) Electron-lattice coupling which has been briefly discussed in the present paper for the specific LO state with $q=\pm\pi$ in the purely 1-D model but not for the other $q$ cases as well as in the realistic quasi-1-D model. 
(2) Strong coupling effect which is important especially in the BEC regime where the excitons are preformed far above $T_c$\cite{Seki2014}. 
(3) Nonmagnetic (normal) impurities which are known to have significant pair-breaking effects for excitons similar to the case of magnetic impurities for Cooper pairs in the superconductivity\cite{Zittartz1967}. 
In addition, the superconductivity of Ta$_2$NiSe$_5$ under high pressure is also interesting future problem as it is observed around the quantum critical point in which the excitonic order disappears. In fact, our preliminary calculation with the random phase approximation reveals that the superconductivity occurs due to the excitonic fluctuation enhanced towards the excitonic phase boundary. 
Explicit results of the superconductivity as well as the above mentioned effects will be reported in subsequent papers. 

\section*{Acknowledgments}
This work was partially supported by a Grant-in-Aid for Scientific Research from the Ministry of Education, Culture, Sports, Science and Technology. 

\bibliographystyle{jpsj.bst}
\bibliography{68656.bib}
\end{document}